\documentclass[aip,rsi,reprint,graphicx]{revtex4-1}
\usepackage{graphicx}
\usepackage{dcolumn}
\usepackage{bm}
\usepackage{float}
\usepackage{hyperref}
\usepackage{color}
\usepackage{cancel}
\usepackage{bmpsize}
\usepackage{amsmath}
\usepackage{amssymb}
\usepackage{natbib}
\newcommand{\specialcell}[2][c]{%
	\begin{tabular}[#1]{@{}c@{}}#2\end{tabular}}

\begin{document}
\title{Noise spectra in balanced optical detectors based on transimpedance  amplifiers}
\author{A. V. Masalov}
\affiliation{International Center for Quantum Optics and Quantum Technologies (Russian Quantum Center), Skolkovo, 143025 Moscow, Russia}
\affiliation{P. N. Lebedev Physical Institute, Leninsky Prospect 53, 119991 Moscow, Russia}
\affiliation{National Research Nuclear University MEPhI, Kashirskoe shosse 31, 115409 Moscow, Russia}
\author{A. Kuzhamuratov}
\affiliation{International Center for Quantum Optics and Quantum Technologies (Russian Quantum Center), Skolkovo, 143025 Moscow, Russia}
\affiliation{Moscow Institute of Physics and Technology, 141701 Dolgoprudny, Moscow region, Russia}
\author{A. I. Lvovsky}
\affiliation{International Center for Quantum Optics and Quantum Technologies (Russian Quantum Center), Skolkovo, 143025 Moscow, Russia}
\affiliation{P. N. Lebedev Physical Institute, Leninsky Prospect 53, 119991 Moscow, Russia}
\affiliation{Institute for Quantum Science and Technology, University of Calgary, Calgary, Alberta T2N 1N4, Canada}
\affiliation{Institute of Fundamental and Frontier Sciences, University of Electronic Science and Technology, Chengdu, Sichuan 610054, China}
\date{\today}
\begin{abstract}
	We present a thorough theoretical analysis and experimental study of the shot and electronic noise spectra of a balanced optical detector based on an operational amplifier (OA) connected in a transimpedance scheme. We identify and quantify the primary parameters responsible for the limitations of the circuit, in particular the bandwidth and shot-to-electronic noise clearance. We find that the shot noise spectrum can be made consistent with the second order Butterworth filter, while the electronic noise grows linearly with the second power of the frequency. Good agreement between the theory and experiment is observed, however the capacitances of the operational amplifier input and the photodiodes appear significantly higher than those specified in manufacturers’ datasheets. This observation is confirmed by independent tests. 
\end{abstract}

\maketitle
\section{Introduction}
Balanced homodyne detection (BHD) is a primary tool of quantum optics. Invented in 1983 by Yuen and Chan\cite{1}, it enables direct measurement of phase-dependent field quadratures of electromagnetic modes. This permits complete characterization of quantum states of light in these modes\cite{2,3}.

In balanced homodyne detection, the quantum field to be measured is brought into interference with a strong coherent laser field – the local oscillator – on a symmetric beam splitter. The fields emerging in the two output channels of the beam splitter are directed onto two high-efficiency photodiodes, whose photocurrents are then subtracted. The subtracted photocurrent is proportional to the quadrature of the field of interest, with the corresponding phase being determined by the optical phase of the local oscillator.

The final detection and subtraction of the optical signal is a delicate task that is often implemented by a dedicated electronic circuit known as the balanced detector. The complexity of this task arises from the subtraction signal being much weaker than that associated with the macroscopic fields incident on each photodiode. Therefore the balanced detector circuit must feature a high common-mode rejection ratio as well as extremely low-noise amplification.

\begin{figure}[t]
	\includegraphics[width=2.3in]{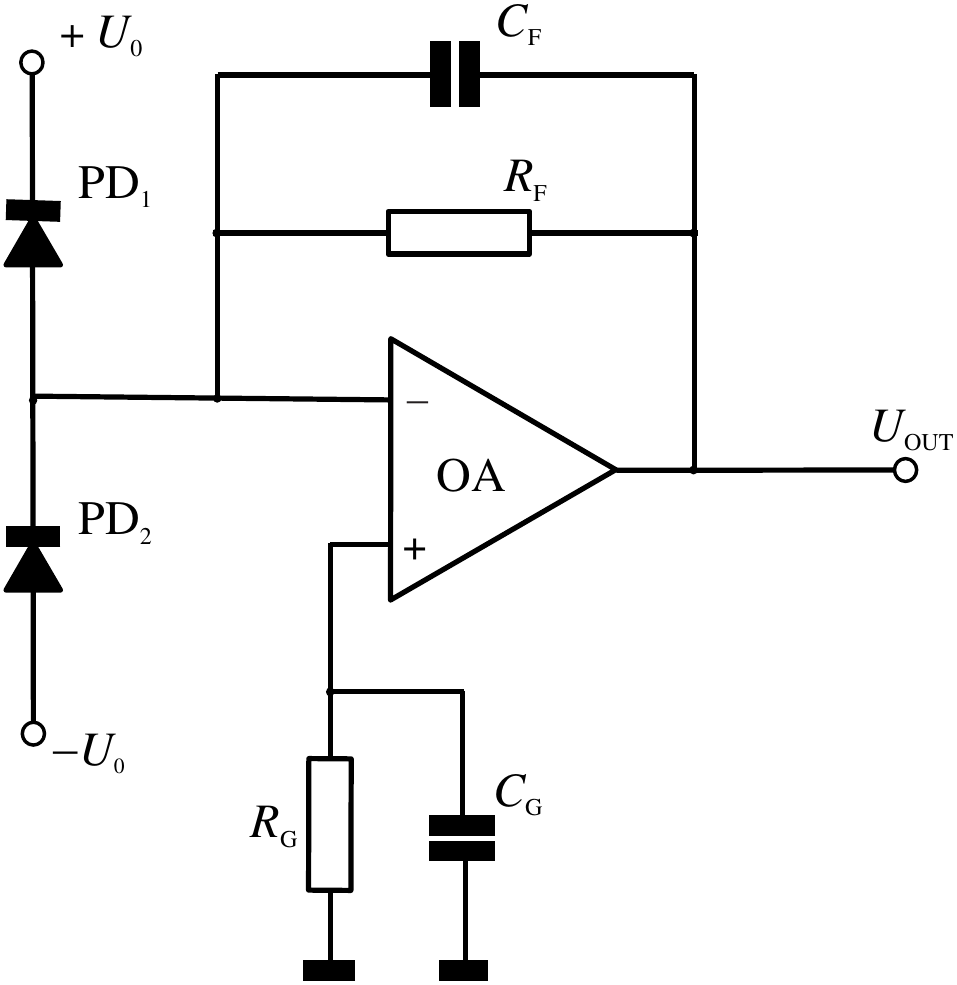}
	\caption{ Scheme of balanced detector circuit with transimpedance operational amplifier.}
	\label{f1}
\end{figure}

In a typical balanced detector circuit, the two photodiodes are connected in series, which permits direct physical subtraction of their photocurrents at their connection point (Fig.~\ref{f1}). The subtraction photocurrent is subjected to amplification. The latter, most typically, occurs in an OA set up in a transimpedance scheme, in which the input weak current is converted into a voltage output\cite{4,5,6,7,8,9,10,11,12}.
The primary characteristics of the amplification circuit are the bandwidth and the electronic noise. The benchmark of the latter is the so-called shot-to-electronic noise clearance, i.e. the ratio of the quantum noise corresponding to the vacuum input field (measured in the presence of the local oscillator whose strength is set to the maximum possible level enabling stable operation) and the electronic noise of the amplifier (measured with the local oscillator field blocked). In many modern balanced detectors, this ratio is on a scale of $10-20$ dB in the working bandwidth of the detector, corresponding to a noise-defined quantum efficiency of $90-99\%$\cite{13}.

While the research on constructing and improving balanced detectors for quantum optics applications is making rapid progress, there is no thorough understanding in the community regarding the origin of the electronic noise; the desired bandwidth and clearance characteristics are often achieved by means of trial and error. In this paper, we address this gap in understanding. We show how the noise characteristics of a transimpedance-based balanced detector can be quantitatively calculated, and corroborate our calculations by experimental results.

\section{Theory} 
\noindent \textbf{A. Amplification spectrum.} A typical transimpedance-based BHD circuit is given in Fig.~\ref{f1}. The bias voltage of photodiodes $ \pm U_{\text{0}}$ reduces their intrinsic capacitances to increase the amplification bandwidth. The capacitance $C_{\text{F}}$ in the feedback loop is used to optimize the flatness of amplification spectrum. It is generally preferred to use the inverting OA input due to better noise characteristics. The noninverting input of the OA should be grounded by a resistor $R_{\text{G}}$ (as a rule it is a variable resistor with shunt capacitance $C_\text{G}$) to eliminate a possible DC offset of the output.

In the Appendix, we derive the expressions for the amplification and electronic noise spectra of the circuit. Neglecting the electronic noise, the amplification spectrum defined as the ratio of the output voltage $U_{\text{OUT}}(f)$ to the input current $I(f)$  at a given frequency $f$ has the form
\begin{equation}
\label{eq1}
\frac{U_{\text{OUT}}(f)}{I(f)}=G(f)R_{F}=\frac{R_{\text{F}}}{1+jp\frac{f}{f^*}-\frac{f^2}{{f^{*}}^2}},
\end{equation}
where $f^*$ and $p$ are two cumulative parameters dependent on the values of all circuit elements (see Appendix):
\begin{equation}
\label{eq2}
f^*=\sqrt{\frac{A_0f_0}{2\pi R_{\text{F}}(2C_{\text{PD}}+C_{\text{F}}+C_{\text{A1}})}}
\end{equation}
has the dimension of frequency and
\begin{equation}
\label{eq3}
p=\left(2\pi R_\text{F} C_\text{F}+\frac{1}{A_0 f_0}\right)f^{*}
\end{equation}
is dimensionless. Here $A_0f_0$ – is the gain-bandwidth product of the OA while $C_{\text{PD}}$ and $C_{\text{A1}}$ are the inherent capacitances of the photodiodes and the OA, respectively, including the parasitic and wiring contribution.

The absolute value of the amplifier gain is
 \begin{equation}
 \label{eq4}
 \left|G(f)\right|^2=\frac{1}{1+(p^2-2)\frac{f^2}{{f^{*}}^2}+\left(\frac{f^2}{{f^{*}}^2}\right)^2}.
 \end{equation}
We can see that, for $p=\sqrt{2}$, the circuit is equivalent to a second-order Butterworth filter, and is characterized by a flat spectrum within its bandwidth. The latter is quantified by the 3-dB cut-off point which, in the case of $p=\sqrt{2}$, becomes equal to $f^*$. In order to achieve this condition, one should choose the feedback capacitance

\begin{equation}
\label{eq5}	
C_{\text{F}}\simeq\sqrt{\frac{2C_{\text{PD}}+C_{\text{A1}}}{\pi A_{0}f_{0}R_{\text{F}}}},
\end{equation}
where we assumed  $C_\text{F}\ll C_{\text{PD}},C_{\text{A1}},(2\pi A_0f_0 R_{\text{F}})^{-1}$.
If a coherent light field (local oscillator) of power $P/2$ is incident on each photodiode, the current signal at the OA input is shot noise with uniform (white) spectral power density 
$\langle I^2 \rangle/\Delta f=2eI_0$, where $I_0=\frac{\eta e}{\hbar \omega} P$ is the total DC photocurrent from both photodiodes. In this case the output voltage noise depends on the frequency according to
\begin{align}\langle U_{\text{OUT}}^2(f)\rangle-\langle U_{\text{e}}^2(f)\rangle&=R_{\text{F}}^2\left|G(f)\right|^2 \langle I^2 \rangle\nonumber\\
&\hspace{-2cm}=\frac{2\eta e^2 P}{\hbar \omega}\Delta f\frac{R_\text{F}^2}{\left(1-\frac{f^2}{{f^*}^2}\right)^2+p^2\left(\frac{f}{f^*}\right)^2},\label{eq6}
\end{align}
where $\langle U_{\text{e}}^2(f)\rangle$ is the electronic noise contribution. The experimental data on the output voltage noise density $\langle U_{\text{OUT}}^2(f)\rangle$ can be fitted to Eq.~\ref{eq6} to give the values of $f^*$ and $p$, and then the capacitance $ 2C_{\text{PD}}+C_{\text{F}}+C_{\text{A1}}$ can be calculated. This value is especially important because of uncertainties of the technical data on $C_{\text{PD}}$ and $C_{\text{A1}}$, as well as the wiring contributions.

\noindent \textbf{B. Electronic noise.}
The electronic noise of the transimpedance amplifier, observed in the absence of input optical signal, originates from inner chip elements of the OA as well as the feedback resistor. This noise is modeled by voltage and current supplies present at the OA input, with the values  $\langle u^2_\text{A}\rangle$ and $\langle i^2_{\text{-}}\rangle$ , respectively.
These values are normally provided in the OA datasheet. An additional contribution is the Nyquist noise associated with the feedback resistor. The noise associated with the dark currents of p-i-n photodiodes is negligibly small.

A derivation of the electronic noise spectral density is given in the Appendix. Relative to the input (i.e. prior to the amplification by the OA) it is
\begin{equation}
\label{eq7}
\frac{	\langle U_{\text{e}}^2(f) \rangle}{\left|G(f)\right|^2R_{\text{F}}^2}=(A+Bf^2) \Delta f,
\end{equation}
where
\begin{equation}
\label{eq8}
A=\frac{\langle u^2_\text{A}\rangle}{R_{\text{F}}^2}+\langle i^2_-\rangle+\frac{4kT}{R_{\text{F}}},
\end{equation}
and 
\begin{equation}
\label{eq9}
 B=[2\pi(2C_{\text{PD}}+C_{\text{F}}+C_{\text{A1}})]^2\langle u^2_\text{A}\rangle=\left(\frac{A_0 f_0}{R_\text{F} {f^*}^2}\right)^2\langle u^2_\text{A}\rangle.
\end{equation}	
\indent We see that the Nyquist and current noise are white while the voltage noise is linearly dependent on the square of the frequency. This is because the noise input voltage, applied to the impedance $[2\pi j f (2C_{\text{PD}}+C_{\text{F}}+C_{\text{A1}})]^{-1}$ associated with the capacitances, gives rise to a current noise that is proportional to the frequency. Estimating $A$ and $B$ by fitting the experimental noise data gives an estimation of the OA noise characteristics $\langle u^2_\text{A}\rangle$ and $ \langle i^2_-\rangle$, which may differ from the datasheet values.

\section{Experiment} 
\begin{figure}[t]
	\includegraphics[width=3.4in]{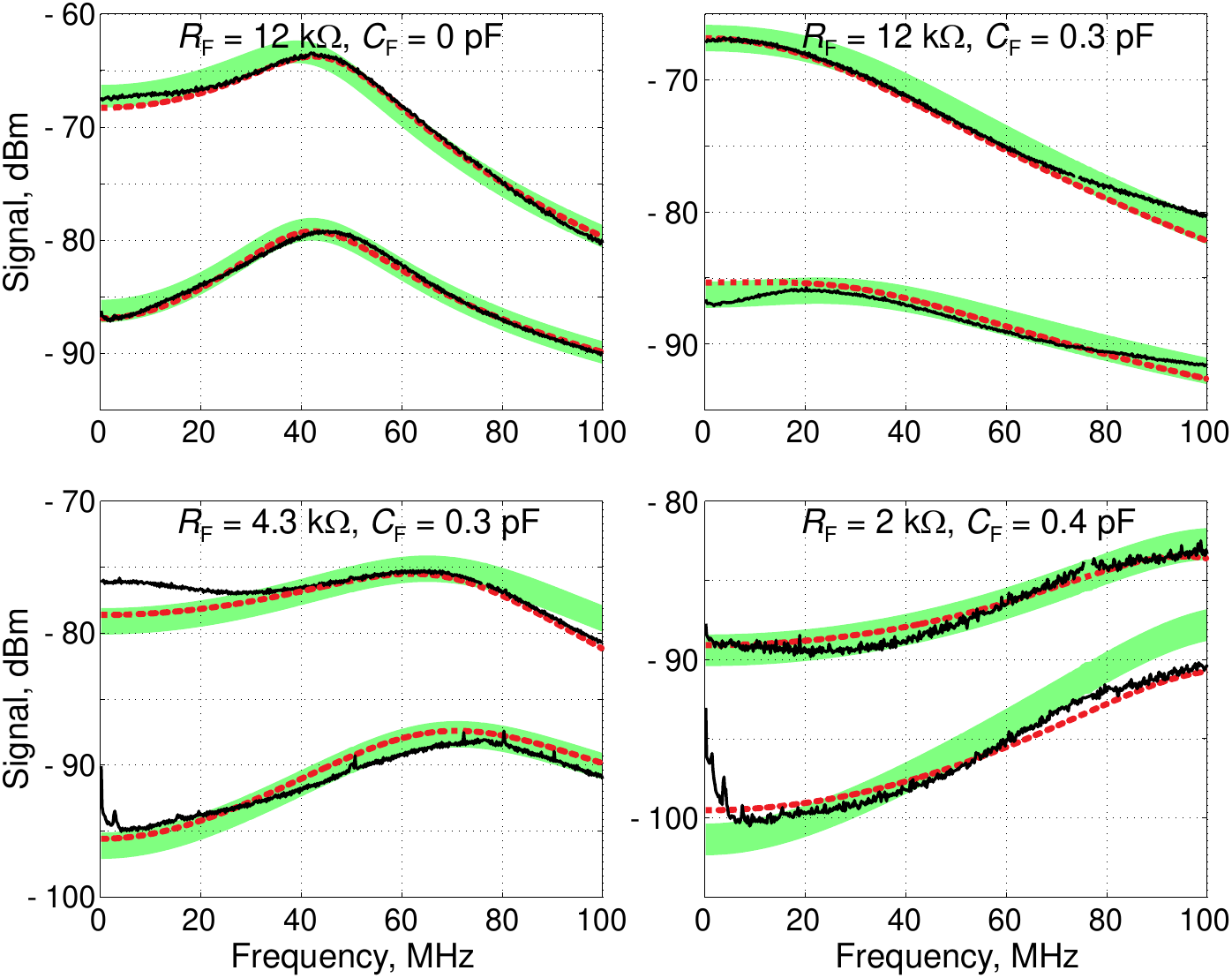}
	\caption{Experimental data on the balanced detector output spectrum for several settings of the feedback resistance and capacitance. The top black curve in each graph corresponds to the input optical power of 10.6 mW (2.7 mW for $R_{\text{F}}$ = 2 k$\Omega$). The bottom black curve corresponds to the electronic noise. Spectra for four different sets of the feedback resistor and capacitance are displayed. The sharp peaks at the laser repetition rate 76 MHz, which emerged due to non-ideal balance of the photocurrents, were manually removed from the observed spectra. The red dashed lines show the best fits with the parameters in Tables \ref{t1} and \ref{t2}. The green  $\pm$ 1 dB corridors show fits with the same capacitance parameter for all four data sets.}
	\label{f2}
\end{figure}

Our detector largely followed the circuit of Kumar {\it et al.}\cite{4}, with an improved circuit layout to reduce the spurious capacitances and inductances. The OA was OPA847 (Texas Instruments) in accordance with the scheme in Fig.~\ref{f2}, where $R_\text{G}=R_\text{F},C_\text{G}=0,1$  $\mu$F, while $R_\text{F}$ and $C_{\text{F}}$ were varied; two photodiodes S5972 (Si, Hamamatsu) were used. To check the validity of our noise model and to estimate the unknown parameters of the circuit elements we have measured the detector’s output power spectrum both in presence and absence of the optical (local oscillator) field. The latter was generated by a Ti:Sapphire laser emitting a train of 1.7-ps pulses at 780 nm (repetition rate 76 MHz). Examples of the detector’s output noise spectra measured by an Anritsu MS2034B spectrum analyzer for different $R_\text{F}$ and $C_\text{F}$ are shown in Fig.~\ref{f1}. We used the light power  = 10.6 mW except for $R_\text{F}$ = 2 k$\Omega$, in which case the power was reduced to 2.7 mW. This is because the OA exhibited a tendency to self-oscillate at higher powers, which is a typical behavior of OAs for low feedback resistances. Electronic noise samples for each parameter sets are also displayed in the figure.

The shot noise spectra were fitted with Eq.~\eqref{eq6}. The fitting parameters were the values $f ^*$ and $p$, and then they were used to extract the values of $2C_{\text{PD}}+C_\text{F}+C_{\text{A1}}$  and $C_\text{F}$. The results are given in Table \ref{t1}.

 Only one dataset ($R_\text{F} = 12$ kOhm, $C_\text{F} = 0.3$ pF) exhibits a behavior consistent with $p>\sqrt{2}$ , i.e. with no maximum at $f > 0$. In other cases, the measured shot noise exhibits a smooth maximum in the range 40-75 MHz, corresponding to $p<\sqrt{2}$ ; the dependence \eqref{eq4} gives a good approximation of the noise spectrum around the maximum towards higher frequencies. For lower frequencies, the shot noise deviates upwards from the theoretical fit, particularly for $R_\text{F} = 4.3$ k$\Omega$. As a result, the setting with this feedback resistor yields a reasonable flat spectrum which is suitable for time-domain quantum measurements.

There is good consistency between the values of $2C_{\text{PD}}+C_\text{F}+C_{\text{A1}}$ estimated from the data for different $R_\text{F}$: $2C_{\text{PD}}+C_\text{F}+C_{\text{A1}}=23\div29$ pF. To illustrate this, we also approximated all our data sets (with different $R_\text{F}$ and $C_\text{F}$) with a unified value $2C_{\text{PD}}+C_{\text{A1}}=26$ pF (including parasitic and wiring contributions) and actual $C_\text{F}$ values: $0, 0.3, 0.3$ and $0.4 $ pF with a parasitic addition of 0.2, 0.2, 0.1 and 0 pF (respectively). These approximations are shown by green corridors in Fig.~\ref{f2} and agree with the data within a $\pm$ 1 dB range.
\begin{table}[H]
	\caption{Amplifier parameters and overall capacitances, best fit to the experimental data.}
	\label{t1}
\begin{center}
	\begin{tabular}{|c|c|c|c|}
		\hline
	\specialcell{\\ \\ } &$f^*$, MHz &$p$ & $2C_{\text{PD}}+C_\text{F}+C_{\text{A1}}$, pF \\
		\hline 
	\specialcell{$R_\text{F}=12$ kOm \\ $C_{\text{F}}=0$ pF}& 47.3$\pm$ 3&0,63$\pm$0.05 &23.1$\pm2$	\\
		\hline
	  \specialcell{$R_\text{F}=12$ kOm\\   $C_{\text{F}}=0,3$ pF}& 44.5$\pm$ 3&1.9$\pm$0.1 &26.1$\pm2$	\\
	\hline
	 \specialcell{$R_\text{F}=4.3$ kOm\\ $C_{\text{F}}=0.3$ pF} & 73.3$\pm$ 4&0,75$\pm$0.05 &26.8$\pm2$	\\
		\hline
		\specialcell{$R_\text{F}=2$ kOm \\ $C_{\text{F}}=0.4$ pF}& 103$\pm$ 5&0,54$\pm$0.05 &29.4$\pm3$	\\
		\hline	
	\end{tabular}
\end{center}
\end{table}
The value of $2C_{\text{PD}}+C_{\text{A1}}=26$ pF is much higher than one would expect from the technical data on $C_{\text{PD}}$ and $C_{\text{A1}}$. According to the data sheet of the S5972 photodiode\cite{14}, $C_{\text{PD}}\sim$ 3 pF(at $U_{0}=10$ V), while for OPA847\cite{15}, $C_{\text{A1}}\sim$ 2 pF. Assuming the parasitic and wiring capacitance to be about 2$\div$3 pF, the deficit of about 15 pF is present. We conducted further research, described below, and found that this deficit should be attributed to both $C_{\text{PD}}$ and $C_{\text{A1}}$.

\begin{figure}[t]
	\includegraphics[width=3.4in]{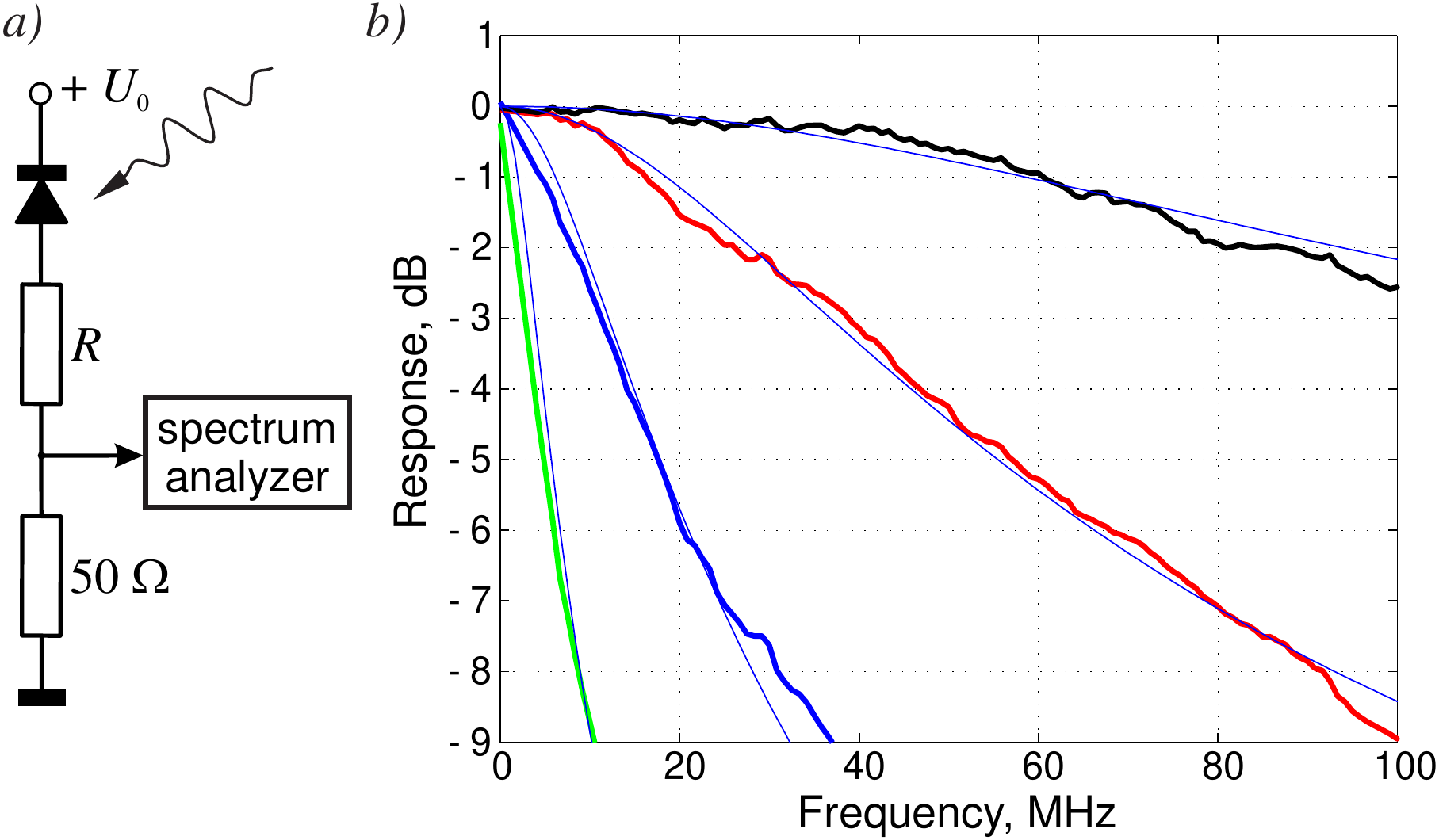}
	\caption{Measuring the capacitance of the S5972 photodiode by observing its shot noise spectrum. a) Experimental circuit; b) Measured spectra for different load resistors: 160 $\Omega$ (black), 510 $\Omega$ (red), 1.6 k$\Omega$ (blue), 5.1 k$\Omega$ (green). Thin lines show theoretical fits with $C_{\text{PD}} = 8$ pF.}
	\label{f3}
\end{figure}

The Texas Instruments application notes on OPA847\cite{15} provides amplification spectra for the OA connected to a diode in a transimpedance scheme similar to ours. We demonstrate in the Appendix that these spectra can be fit with $C_{\text{A1}} = 10$ pF but not $C_{\text{A1}} = 2$ pF). Furthermore, observations of the excessive input capacitance of OPA847 have been reported by other users of this OA on the Texas Instruments community internet forum\cite{16}.
In order to measure the capacitance of the S5972 photodiode, we constructed a circuit shown in Fig.~\ref{f3}(a). The photodiode was illuminated with a laser diode. The shot noise generated by the photodiode was subjected to low pass filtering by an integrating circuit formed by the intrinsic capacitance of the photodiode and a resistor. The spectrum of that shot noise was then measured by the spectrum analyzer. The resulting data [Fig.~\ref{f3}(b)]  showed a good fit with $C_{\text{PD}} = (8\pm 1)$ pF.

Thus, we can conclude that the total value $C_{\text{PD}}+C_{\text{A1}}= 26$ pF resulted in our measurements consists of a photodiode contribution  $2C_{\text{PD}}=16$ pF and that of the amplifier $C_{\text{A1}}= 10$ pF (both including parasitic and wiring contributions).

\begin{figure}[t]
	\includegraphics[width=3.4in]{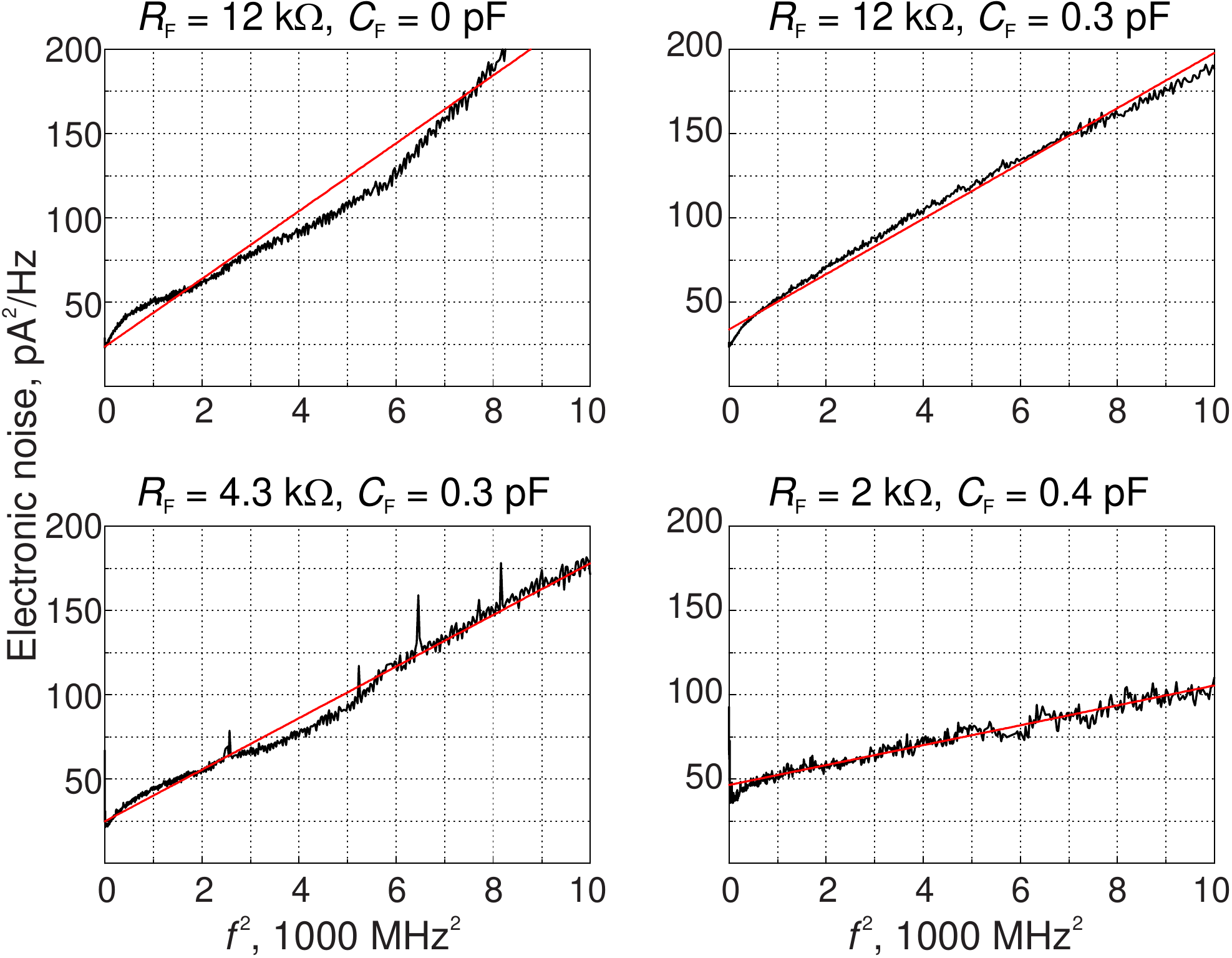}
	\caption{Electronic noise relative to the OA input for four settings of the feedback resistance and capacitance. Expreimental data and theoretical fit are displayed in each graph.}
	\label{f4}
\end{figure}

The parameters obtained in fitting the shot noise data were used in the quantitative analysis of the electronic noise in order to relate it to the amplifier input. The experimental electronic noise spectra divided by corresponding $\left|G(f)\right|^2$ are shown on Fig.~\ref{f4}. In all cases, the resulting spectra are well approximated by linear dependence on the square of the frequency in accordance with \ref{eq7}. These linear dependences allowed us to estimate  $\langle u^2_\text{A}\rangle$ and $\langle i^2_-\rangle$  (Table \ref{t2}). The data for $R_\text{F} = 4.3 \div 12$ k$\Omega$ demonstrate good consistency with each other, while the plot for $R_\text{F} = 2$ k$\Omega$ exhibited a significantly lower slope, corresponding to a lower $\langle u^2_\text{A}\rangle$. This discrepancy is ascribed to the mentioned above instability of the OA observed for low feedback resistances.
\begin{table}[H]
	\caption{Amplifier noise parameters, best fit to the experimental data.}
	\label{t2}
	\begin{center}
		\begin{tabular}{|c|c|c|}
			\hline
			&$\sqrt{\langle u^2_\text{A}\rangle}$,nV/Hz$^{1/2}$ & $ \sqrt{\langle i^2_-\rangle} $, pA/Hz$^{1/2}$ \\
			\hline 
		\specialcell{	$R_\text{F}=12$ kOm \\ $C_{\text{F}}=0$ pF}&0,97 $\pm$0,2&4,7$\pm$0,5	\\
				
			\hline
		\specialcell{$R_\text{F}=12$ kOm \\ $C_{\text{F}}=0,3$ pF}&0,78 $\pm$0,2&5,6$\pm0,5$	\\
			
			\hline
		\specialcell{$R_\text{F}=4.3$ kOm\\ $C_{\text{F}}=0.3$ pF}&0,73 $\pm$0,2 &4,6$\pm0,5$	\\
			
			\hline
			\specialcell{$R_\text{F}=2$ kOm\\ $C_{\text{F}}=0.4$ pF}&0,42 $\pm$0,2 &6,2$\pm$ 0,5	\\
		
			\hline	
		\specialcell{ \\ Unified set \\ \\}&0,85 &5,1	\\
			\hline	
		\end{tabular}
	\end{center}
\end{table}

Our data on voltage noise are in good agreement with those in the data sheet of OPA847\cite{15},  $\sqrt{ \langle u^2_\text{A}\rangle} = (0.85 \div 0.92)$ nV/Hz$^{1/2}$, while the current noise level is one and a half larger than that in the data sheet\cite{15},  $ \sqrt{\langle i^2_-\rangle} = (2.7 \div 3.5)$ pA/Hz$^{1/2}$. Similarly to the shot noise case, we approximated the electronic noise data with a unified parameter set and obtained a consistent agreement within $\pm1$ dB except for $R_\text{F} = 2$ k$\Omega$.

We summarize the contributions to the electronic noise, related to the input of OPA847, below. The range of $R_\text{F}$ is assumed to be between 2 and 12 k$\Omega$.
\begin{itemize}
	\item Thermal noise of the feedback resistor:
	 \begin{equation*}
	  \frac{4kT}{R_{\text{F}}}\approx\frac{16[\text{kOhm}\times\text{pA}^2/\text{Hz}]}{R_{\text{F}}}\sim(1.3\div8) \left[\frac{\text{pA}^2}{\text{Hz}}\right],
	  \end{equation*}
	\item Current noise of OPA847: \begin{equation*}		
 \langle i^2_-\rangle\approx20\div30\left[\frac{\text{pA}^2}{\text{Hz}}\right],	
 \end{equation*}
	\item Voltage noise of OPA847:
	\begin{equation*}
	\frac{\langle u^2_\text{A}\rangle}{R_\text{F}^2}\left(1+\left(\frac{A_0f_0}{{f^*}^2}\right)^2f^2\right)	\sim(0.006\div0.2)\left[\frac{\text{pA}^2}{\text{Hz}}.\frac{1}{\text{MHz}^2}\right]f^2.
	\end{equation*}

\end{itemize}

The experimental data on noise and estimations show that the thermal noise of $R_\text{F}$ has the lowest contribution to the overall noise, while the current and thermal OA noise contribute substantially in the region of low and high frequencies, respectively.
\section{Discussion} An important parameter of balanced optical detectors is the ratio of the shot to electronic noise levels, also known as the clearance. It is a convenient parameter because it is independent of the amplification spectrum, i.e. remains virtually the same if subsequent amplification stages are applied to the output signal. From Eqs.~\eqref{eq6} and \eqref{eq7}, we find for the clearance
\begin{equation}
\label{eq10}
\frac{ \langle U^2_{\text{OUT}}\rangle}{ \langle U^2_\text{e}\rangle}=2\frac{\eta e^2 P_{\text{max}}}{\hbar\omega}\frac{1}{(A+Bf^2)}+1
\end{equation}
Here $P_{\text{max}}$  is the highest optical power that can be applied to the photodiodes without saturating them, which is on a scale of 10-15 mW in our experiment. According to Eq.~\eqref{eq4}, the clearance value is a frequency dependent parameter. The increase of the electronic noise with the frequency causes the clearance to degrade.

If the OA’s current noise is the primary contribution to the electronic noise at DC (as is the case with our experiment), Eq.~\eqref{eq7} at $f = 0$ simplifies to
\begin{equation}
\label{eq11}
\frac{ \langle U^2_\text{OUT}\rangle}{ \langle U^2_\text{e}\rangle}\approx 2\frac{\eta e^2 P_{\text{max}}}{\hbar\omega}\frac{1}{\langle i^2_-\rangle}+1=\frac{2eI_0}{\langle i^2_-\rangle}+1.
\end{equation}
This corresponds to 19 dB for $P_{\text{max}} = 10.6$ mW assuming $\eta = 0.9$, quite in agreement with Fig.~\ref{f2}. Remarkably, this result is independent of the feedback resistance $R_\text{F}$. On the other hand, increasing this parameter will reduce the detector bandwidth $f^*$ [Eq.~\eqref{eq2}] while keeping the proportionality coefficient $B$ constant. We conclude that there is no benefit in increasing $R_\text{F}$; one should choose the lowest possible value that prevents self-oscillation. In our experiment, this is about 4 k$\Omega$, corresponding to $f^* = 70-100$ MHz bandwidth. If the capacitances of the OA and the photodiodes were consistent with the data sheet values, this parameter would increase by a factor of $\sqrt{26 \text{ pF}/8\text{ pF}}=1.8$. Furthermore, if the intrinsic current noise of the OA were consistent with the data sheet, we would see an additional 3 dB in clearance. Further improvements in the performance of balanced detectors can be achieved by using a field-effect transistor instead of an OA for the first cascade of amplification\cite{17}.

To summarize the paper, we demonstrate that the experimental data on the shot noise spectra in a balanced optical detector are well-fitted by an inverse second-order polynomial function of the  frequency squared, while the experimental data on OPA electronic noise spectra (at input) are well-fitted by a linear dependence on the square of the frequency.   Our results permit one to predict the noise and bandwidth characteristics of a transimpedance-OA-based balanced optical detector provided that the capacitances of the OA and photodiodes, as well as the noise characteristics of the OA, are known.

\appendix
	\section{Gain spectrum calculation.}
	Here we apply the general principles of calculating electronic spectra put forward e.g. in Ref.~\cite{18} to the balanced optical detector (Fig.~\ref{f1}).
	
\begin{figure}[t]
		\includegraphics[width=2.5in]{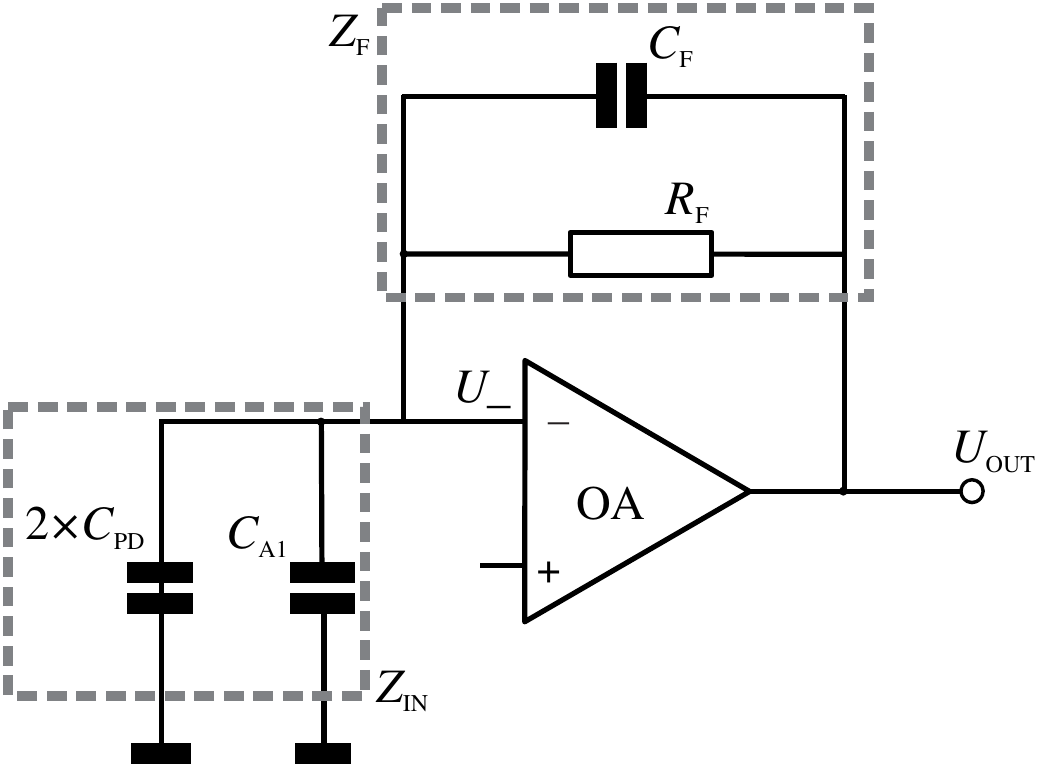}
		\caption{Equivalent scheme of the balanced detector used in the gain spectrum calculation.}
		\label{A1}
	\end{figure}

	The equivalent scheme of a balanced detector used in the gain spectrum calculations is presented in Fig.~\ref{A1}, where the respective intrinsic capacitances $C_{\text{PD}}$ and $C_{\text{A1}}$ of the photodiodes and the inverting input cascade of the OA are taken into account. We can define the cumulative impedance of the elements in the input and feedback parts of the circuit (Fig.~\ref{A1}):
	\begin{equation*}
	\frac{1}{Z_{\text{F}}}=\frac{1}{R_{\text{F}}}+2\pi j f C_{\text{F}} 
	\text{ and } 
	\frac{1}{Z_{\text{IN}}}=2 \pi j f (2C_{\text{PD}}+C_{\text{A1}}).
	\end{equation*}
	The photocurrent signal $I(f)$  is distributed between these parts, while the current into the inverting input of OA is negligible. This results in the equation
	\begin{equation}
	\label{eq12}
	I(f)=\frac{U_-}{Z_{\text{IN}}}+(U_--U_{\text{OUT}})\frac{1}{Z_\text{F}}.
	\end{equation}
	The input and output voltages of the OA are related by $U_{\text{OUT}}=-A(f)U_-$,where 
	$A(f)=A_0/(1+jf/f_0)$ is the intrinsic OA gain ($A_0f_0$ being the gain-bandwidth product of OA). Solving Eq. \ref*{eq12}, we find the frequency response of the transimpedance amplifier:
	\begin{widetext}
	\begin{equation}
	\frac{U_{\text{OUT}}}{I(f)R_{\text{F}}}=-\frac{1}{R_{\text{F}}}\bigg/\left[\frac{1}{Z_{\text{F}}}+\frac{1}{A(f)}\left(\frac{1}{Z_\text{F}}+\frac{1}{Z_{\text{IN}}}\right)\right]
	\label{eq13}
	\approx -1\bigg/\bigg[1+jf\left(2\pi R_\text{F} C_\text{F}+\frac{1}{A_0f_0}\right)
	-\frac{2\pi f^2}{A_0 f_0}R_\text{F}(2C_{\text{PD}}+C_{\text{F}}+C_{\text{A1}})\bigg].
	\end{equation}
	\end{widetext}
	This is equivalent to Eq. \ref{eq1} where the polynomial coefficients are expressed through $f^*$ and $p$. In the approximation, we used the inequality 
	$\frac{1}{Z_\text{F}}\gg\frac{1}{A_0}\left(\frac{1}{Z_\text{F}}+\frac{1}{Z_{\text{IN}}}\right)$ valid for large values $A_0 \sim 10^6$.
	
	Two elements outside the OA – the resistor $R_\text{F}$ and capacitor $C_\text{F}$ of the feedback circuit  – can be varied to affect the frequency response, while other parameters - $C_{\text{PD}}$, gain-bandwidth product and $C_{\text{A1}}$ - are fixed by the choice of the photodiodes and OA. Note that all three capacitors in wideband transimpedance amplifier circuits are typically in the range of several pF and their actual capacitances include wiring addition.
	
	\section{Electronic noise calculation}
	The electronic noise of the transimpedance amplifier originates from inner chip elements of the OA as well as the feedback resistor. We consider the noise sources at the inverting and non-inverting inputs of the OA separately.
	
	\subsection{Inverting input} 
\begin{figure}[t]
		\includegraphics[width=\columnwidth]{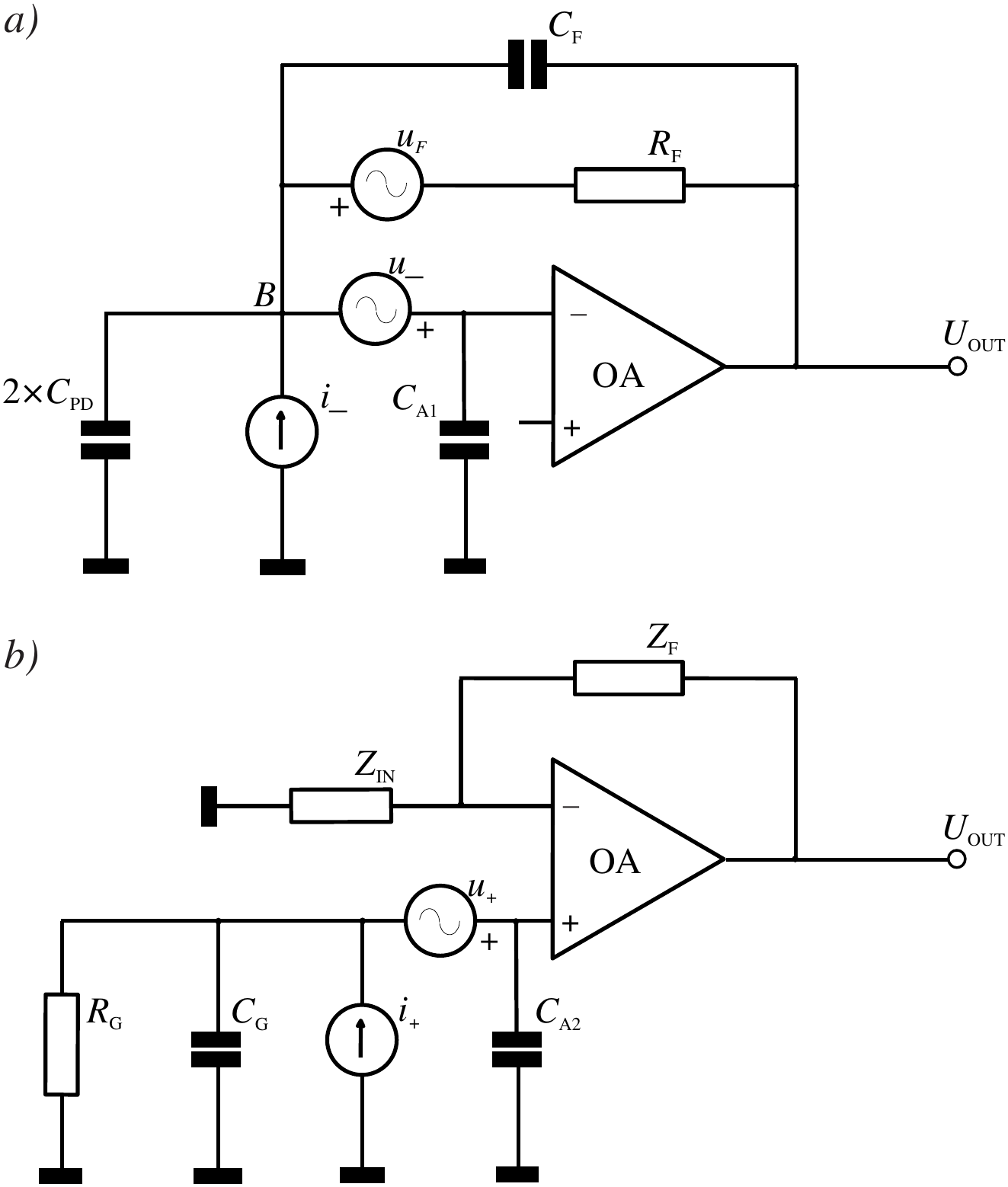}
		\caption{Scheme of a transimpedance amplifier with equivalent noise sources of OA at (a) feedback resistor and inverting input; (b) non-inverting input. }
		\label{A2}
	\end{figure}

Equivalent scheme for calculation of current $i_{-}$ and voltage $u_{-}$ noise contributions of OA is given in Fig.~\hyperref[A2]{6(a)}.
	
	We shall find the contributions of each noise source separately and then combine their squared values.
	
	The contribution of the OA \textit{current noise $i_{-}$} is equivalent in treatment to the signal of the photodiodes:
	$$U_{\text{OUT}}=-R_\text{F}G(f)i_{-}$$
	and 
	\begin{equation}
	\label{eq14}
	\langle U_{\text{OUT}}^2\rangle=R_\text{F}^2\left|G(f)\right|^2 \langle i_{-}^2\rangle.
	\end{equation}
	\indent The OA \textit{voltage noise source $u_{-}$ } produces current through the input elements and the feedback circuit (the input current of the OA is assumed to be zero):
	\begin{equation}
	\label{eq15}
	\frac{U_{\text{OUT}}-U_{\text{B}}}{Z_\text{F}}=2\pi j f (2C_{\text{PD}})U_{\text{B}}+2\pi j f C_{\text{A1}}U_{-},
	\end{equation}
	where $U_{-}$ is the voltage at the inverting input of the OA and $U_{\text{B}}=U_{-}-u_{-}$ is the voltage at point B [Fig.~\ref{A2}(a)]. Since $U_{\text{OUT}}=-A_0u_{-}/(1+jf/f_0)$, this yields
	
	\begin{equation}
	\label{eq16}
	U_{\text{OUT}}=-G(f)\left[1+2\pi j f R_\text{F}(2C_{\text{PD}}+C_\text{F}+C_{\text{A1}})\right]u_{-},
	\end{equation}
	
	This contribution increases linearly with the frequency, and the corresponding power is linearly dependent on the square of the frequency.
	
	The thermal noise $\langle u_F^2\rangle=4kT\Delta f/ R_\text{F}$ of the feedback resistor $R_\text{F}$ sends current through both the input elements and the elements of the feedback circuit:
	\begin{widetext}
	\begin{equation}\label{eq17}
	u_{-}(2\times 2\pi j f C_{\text{PD}}+2\pi j f C_{\text{A1}})
	=(U_{\text{OUT}}-U_{-})2\pi j f C_\text{F}+\frac{U_{\text{OUT}}-U_{-}+u_F}{R_\text{F}},
	\end{equation}
	which yields  $U_{\text{OUT}}=-G(f)u_F$ and hence
	\begin{equation}
	\label{eq18}
	\langle U_{\text{OUT}}^2\rangle=\left|G(f)\right|^2 4kTR_\text{F}\Delta f.
	\end{equation}
	
	\subsection{Noninverting input} The equivalent scheme for the calculation of current $i_{+}$ and voltage $u_{+}$ noise contributions of OA is given in Fig.~\ref{A2}(b). The capacitor $C_\text{G}$ usually has the value $\sim 10^5$ pF, and it shunts the thermal noise of $R_\text{G}$ and the current noise of the OA. Only the voltage noise of the OA significantly affects the output.
	
	The \textit{current noise source} sends the current $i_{+}$ through the input elements and produces the input voltage $U_{+}$:
	\begin{equation}
	\label{eq19}
	U_{+}=\frac{i_+}{\frac{1}{R_\text{G}}+2 \pi j f (C_{\text{G}}+C_{\text{A2}})},
	\end{equation}
	while the inverting input voltage  $U_-$ equals to $U_{\text{OUT}}Z_{\text{IN}}/(Z_{\text{IN}}+Z_{\text{F}})$.Taking into account $U_{\text{OUT}}=A_0(U_+-U_-)/(1+jf/f_0)$, this yields
	\begin{equation}
	\label{eq20}
	U_{\text{OUT}}=R_\text{G} i_+ G(f) \frac{1+2\pi j f R_\text{F}(2C_{\text{PD}}+C_\text{F}+C_{\text{A1}})}{1+2\pi j f R_\text{G} (C_\text{G}+C_{\text{A2}})}.
	\end{equation}
	
	Due to large values of $R_\text{G}$ and $C_\text{G}$, this contribution is noticeable only in a narrow frequency range $\leq1/2\pi R_\text{G} C_\text{G}$, while elsewhere the denominator reduces the output voltage to negligible values. We may conclude that
	\begin{equation}
	\label{eq21}
	\langle U_{\text{OUT}}^2\rangle\ll R_\text{F}^2\left|G(f)\right|^2\langle i_{+}^2\rangle,
	\end{equation}
	and ignore this contribution in further analysis.

The \textit{voltage noise} source ($u_+$) sends current through the input elements and produces the input voltage $U_+$:
	\begin{equation}
	U_{+}=\frac{u_+}{(\frac{1}{R_\text{G}}+2\pi j f C_\text{G})^{-1}+\frac{1}{2\pi j f C_{\text{A2}}}}\frac{1}{2\pi j f C_{\text{A2}}}
	\label{eq22}
	=\frac{1+2\pi j f R_\text{G} C_\text{G}}{1+2\pi j f R_\text{G}(C_\text{G}+C_{\text{A2}})}u_+\approx u_+,
	\end{equation}
	because $C_\text{G}\gg C_{\text{A2}}$. Taking into account $U_-=U_{\text{OUT}}Z_{\text{IN}}/(Z_{\text{IN}}+Z_{\text{F}})$ and $U_{\text{OUT}}=A_O(U_+-U_-)/(1+jf/f_0)$, this yields 
	\begin{equation}
	\label{eq23}
	U_{\text{OUT}}=G(f)\left[1+2\pi j f R_\text{F}(2C_{\text{PD}}+C_{\text{F}}+C_{\text{A1}})\right]u_+
	\end{equation}
	This contribution behaves in exactly by the same manner as that of the OA voltage noise at the inverting input (\ref{eq16}). Thus we may characterize the OA voltage noise by the cumulative value $\langle u^2_\text{A}\rangle=\langle u^2_+\rangle+\langle u^2_-\rangle$ for both inputs.

The \textit{thermal noise of the ground resistor} $R_\text{G}$, when represented by a current source, is equivalent to the current noise of the OA. Thus  we have the equation similar to (\ref{eq20}):
	\begin{equation}
	U_{\text{OUT}}
	=R_\text{G}\sqrt{\frac{4kT\Delta f}{R_\text{G}}}G(f)\frac{1+2\pi j f R_\text{F}(2C_{\text{PD}}+C_\text{F}+C_{\text{A1}})}{1+2\pi j f R_\text{G}(C_\text{G}+C_{\text{A2}})}
	\label{eq24}
	\approx R_\text{F}\sqrt{\frac{4kT\Delta f}{R_\text{G}}}G(f)\frac{2C_{\text{PD}}+C_\text{F}+C_{\text{A1}}}{C_\text{G}}
	\end{equation}
	At $R_\text{G}\sim (3\div 10)$ k$\Omega$ its thermal noise has an order of magnitude $(1\div2) $ pA/Hz$^{1/2}$, which is close to the current noise of a low-noise OA. Thus, due to the factor  $(2C_{\text{PD}}+C_{\text{F}}+C_{\text{A1}})/C_\text{G}\ll 1$, this contribution can also be neglected.
	\subsection{Total noise.} The sum of all noise contributions has the form:
	\begin{equation}
	\label{eq25}
	\frac{ \langle U_{\text{OUT}}^2\rangle}{\left|G(f)\right|^2}=\left[1+4\pi^2R_\text{F}^2(2C_{\text{PD}}+C_{\text{F}}+C_{\text{A1}})^2f^2\right]\langle u_\text{A}^2\rangle+R_\text{F}^2\langle i_-^2\rangle+4kTR_\text{F}\Delta f 
	\end{equation}
	Where the right-hand side represents the noise relative to the OA input.
	\end{widetext}
	\section{Simulating OPA847 data.}
\begin{figure}[t]
		\includegraphics[width=3.4in]{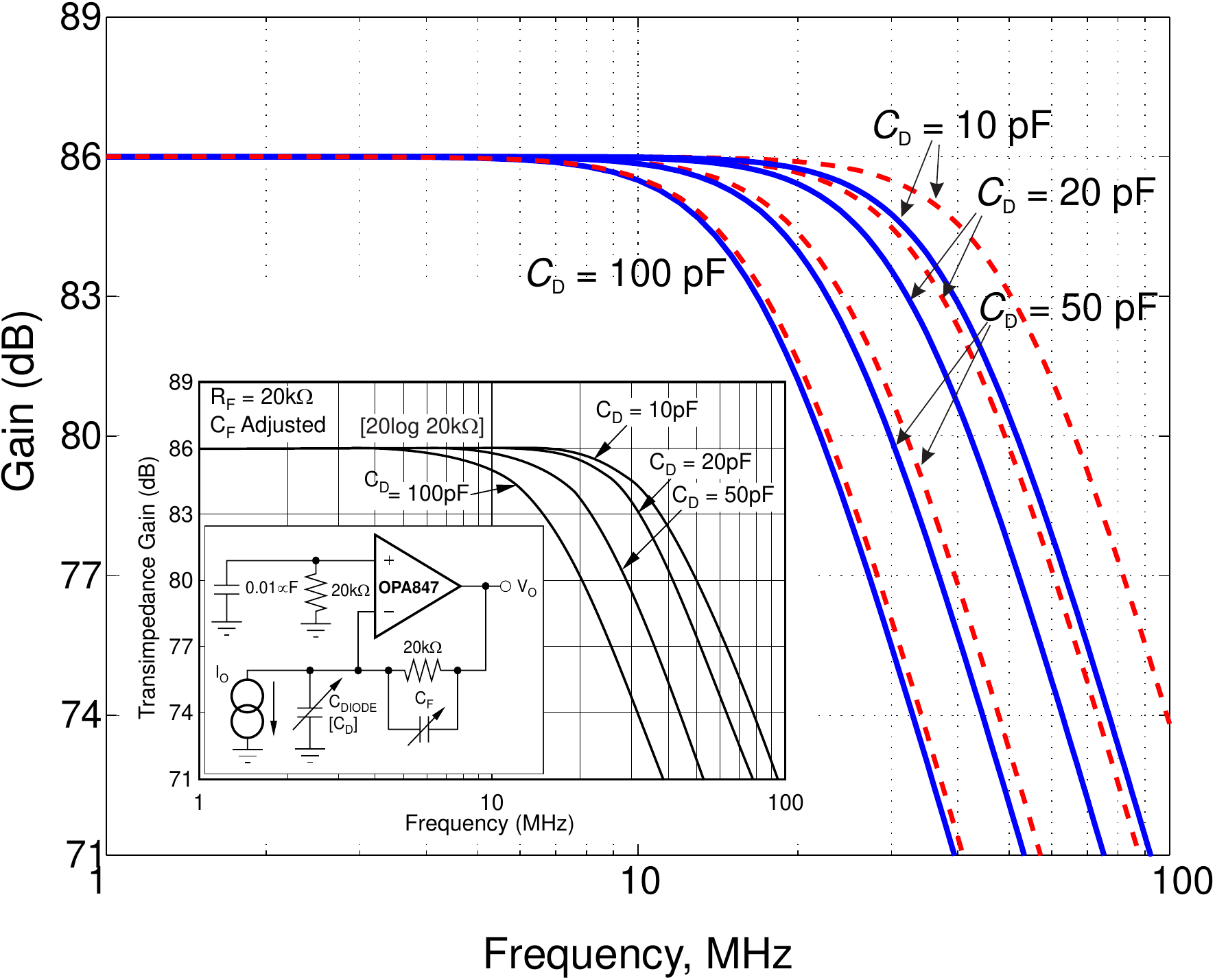}
		\caption{Computational simulation of OPA847 transimpedance frequency response to a signal from a photodiode with the photodiode capacitance $C_\text{D} = 100$ pF, 50 pF, 20 pF and 10 pF with $C_{\text{A1}} = 10$ pF (blue curves), $C_{\text{A1}} = 2$ pF (red dashed curves). The feedback capacitance $C_\text{F}$ is adjusted for $p=\sqrt{2}$ at each $C_\text{D}$. Inset: reproduction of the figure from OPA847 application notes\cite{15}.}
		\label{A3}
	\end{figure}

	We made computational simulation of the photodiode transimpedance frequency response spectra shown on page 8 of the Texas Instruments OPA847 application notes\cite{15} by Eq.~\eqref{eq4} with $R_\text{F} = 20$ k$\Omega$ and $C_{\text{F}}$ adjusted for $p=\sqrt{2}$ . The results are presented in Fig.~\ref{A3}. The calculated spectra with $C_{\text{A1}} = 10$ pF perfectly coincide with those in the application notes\cite{15}, while spectra with $C_{\text{A1}} = 2$ pF are far from agreement, particularly for low values of the photodiode capacitance $C_\text{D}$.

\end{document}